\documentclass[aps,twocolumn,pre,showpacs]{revtex4}
\usepackage{amssymb}
\usepackage{graphicx}

\begin{document}

\title{Lift and drag forces on an inclined plow moving over a granular surface}


\author{Baptiste Percier}
\affiliation{Universit\'e de Lyon, Laboratoire de Physique, Ecole Normale Sup\'erieure de Lyon, \\
CNRS UMR 5672, 46 All\'ee d'Italie, 69364 Lyon cedex 07, France.}
\author{Sebastien Manneville}
\affiliation{Universit\'e de Lyon, Laboratoire de Physique, Ecole Normale Sup\'erieure de Lyon, \\
CNRS UMR 5672, 46 All\'ee d'Italie, 69364 Lyon cedex 07, France.}

\author{Jim N. McElwaine}
\affiliation{DAMTP, University of Cambridge, Wilberforce Rd., CB3 0WA Cambridge, U.K.}
\author{Stephen W. Morris}
\affiliation{Department of Physics, University of Toronto, 60 St. George St., Toronto, Ontario, Canada, M5S 1A7}

\author{Nicolas Taberlet}
\affiliation{Universit\'e de Lyon, Laboratoire de Physique, Ecole Normale Sup\'erieure de Lyon, \\
CNRS UMR 5672, 46 All\'ee d'Italie, 69364 Lyon cedex 07, France. }

\date{\today}

\begin{abstract}

We studied the drag and lift forces acting on an inclined plate while
it is dragged on the surface of a granular media, both in experiment
and numerical simulation. In particular, we investigated the influence
of the horizontal velocity of the plate and its angle of attack. We
show that a steady wedge of grains is moved in front of the plow and
that the lift and drag forces are proportional to the weight of this
wedge. These constants of proportionality vary with the angle of attack but not (or only weakly) on the velocity.
We found a universal effective friction law which accounts for the dependence on all the above-mentioned parameters.
The stress and velocity fields are calculated from the numerical simulations and
show the existence of a shear band under the wedge and that the
pressure is non-hydrostatic. The strongest gradients in stress and
shear occur at the base of the plow where the dissipation rate is
therefore highest.

\end{abstract}
\maketitle


The forces required to disturb the surface of soil have been an important concern of humankind since the invention of the plow, the principal animal-powered tool for this task, about 6\,000\,y ago~\cite{history_plow}.
In this paper we consider the forces on the simplest possible plow, a flat blade inclined in the direction of motion, interacting with the simplest possible soil, a non-cohesive granular material.  Remarkably, this ancient problem has recently received significant attention~\cite{Gravish2010,Ding2011} because of renewed interest in the complex and poorly understood rheology of dry granular materials~\cite{granular_review}.

A simple inclined blade has also been studied as a surrogate for the more complicated situation of a rolling wheel moving over a granular roadbed~\cite{mather62,mather63,heath80,mays00,both01,taberlet2007,Bitbol2009,Hewitt2011}.  Both plows and rolling wheels exhibit an oscillatory instability which produces a spontaneous rippling of the roadbed, leading to a condition known as {\it washboard} or {\it corrugated} road.  Washboard ripples bedevil drivers on unpaved roads worldwide and their mitigation is a serious engineering challenge~\cite{riley,grau,shoop06}.  The modern framework of nonlinear pattern formation~\cite{cross} gives new insight into the formation of washboard ripples~\cite{taberlet2007,Bitbol2009,Hewitt2011}.
A key feature of the washboard instability is the existence of a critical speed $v_c$ below which the flat roadbed is stable.  It has been shown that neither a spring and dashpot suspension, nor compaction of the roadbed are essential to the existence of the instability~\cite{taberlet2007}.  For a wide plow, the problem can be reasonably studied in a 2D vertical plane.
   Dimensional analysis arguments suggest~\cite{taberlet2007,Bitbol2009} that the critical speed for the instability scales as $v_c \sim (mg^2/\rho w)^{1/4}$, where $g$ is the acceleration due to gravity, $m$ is the mass of the plow, $w$ is its width and $\rho$ is the density of the granular material~\cite{Bitbol2009}.


  In this paper, we examine the case of a fixed plow using a combination of experiment and molecular dynamics simulation.
 An understanding of this basic state is a pre-requisite to the elucidation of its subsequent instability to form a washboard pattern.
To do this, we must account for the lift and drag forces experienced by the plow as a function of its speed $v$ and its vertical position $y$, relative to the position of the undisturbed surface $y_0$.  These forces are related to their familiar hydrodynamic equivalents, but, as we will show, a  straightforward fluid mechanical analogy is not particularly helpful. 

The lift and drag forces acting on a totally immersed intruder in a granular medium have been studied for more than 30 years~\cite{Wieghardt1975, Chehata2003,Wassgren2003,Boudet2006,recent_drag}. In a recent paper on immersed intruders~\cite{Ding2011}, it was observed that the lift and drag forces exhibit a strong correlation,  indicating that they scale similarly with system parameters.  Gravish {\it et al.}~\cite{Gravish2010} considered the drag on a vertical blade which plowed the free surface of a granular medium and found oscillatory flow in certain regimes. 
This paper concerns the similar case of a fixed inclined blade for which this type of oscillation is not observed. Our main result is that the plowed material behaves as a solid block sliding over the granular bed. By focusing on the basic state of a flat bed, this study opens the way to a better understanding of the washboard instability. 

\begin{figure}
\includegraphics[width=5cm]{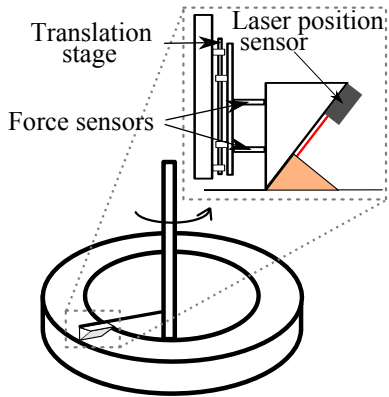}
\caption{A schematic view of the experiment. The system consists of a circular track which is 5-m long, 25-cm wide and 25-cm high filled with sand. A plow is moved over the sand bed at constant (but variable) horizontal velocity and vertical position.}
\label{apparatus}
\end{figure}

This paper is organized as follows: in Sec.~\ref{experiment} and  Sec.~\ref{DEM_simulation} we describe the experimental apparatus and the numerical simulations. The results of both of these  are presented in Sec.~\ref{results}.  
Sec.~\ref{discussion} contains a general discussion, while  
Sec.~\ref{conclusion} presents our conclusions.


\section{Experiment \label{experiment}}


The experimental apparatus, shown schematically in Fig.~\ref{apparatus}, consisted of a  circular track which is 25~cm high and 25~cm wide. It is filled with sand of typical grain size 300~$\pm$~100~$\rm{\mu}$m.  The circumference of the track was $L = 5$~m.

A plow consisting of a flat, inclined blade was moved around the track by a rotating arm.  It was held at a fixed, but adjustable, vertical position $y$, relative to the position $y_0$ of the undisturbed bed.  The plow was a 15 cm wide PVC plate and its angle of attack $\alpha$ with respect to the horizontal could be varied.
%
%
The plow blade was rigidly attached to a translation stage, which allows its vertical position $y$ to be adjusted to within a precision of 5 $\rm{\mu}$m. The speed of the plow over the roadbed can be varied from 0.1 to 1.5~$\rm{m .s^{-1}}$, which covers the range of speed where the washboard instability occurred in previous experiments~\cite{both01}.  However, in the fixed plow experiments we discuss here, no washboard instability occurs and the roadbed is always smoothed by the motion of the plow. These values of the speed are high enough to produce a continuous flow regime and low enough to avoid a gaseous regime.

\begin{figure}
\includegraphics[width=8cm]{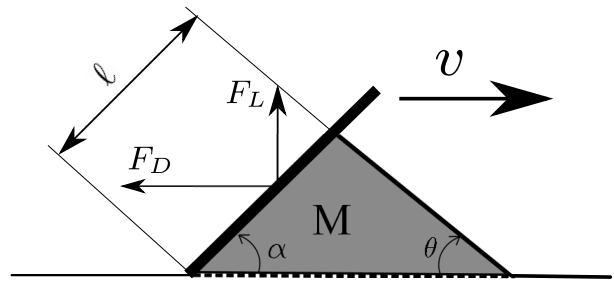}
\caption{A schematic view of the forces on the plow and of the geometry of the mound of plowed sand. The drag force $F_D$ is defined to be the horizontal component of the force, while the lift $F_L$ is the vertical one. $M$ is the total mass of the mound of plowed sand, which varies with the speed of the plow $v$.}
\label{mound}
\end{figure}

The contact forces acting between the plow and the sand were measured by two force transducers (Testwell KD40S) operating in parallel. Using two transducers reduces the torque acting on them and increases the stiffness of the plow support system. The force transducers provide a voltage proportional to the stress exerted along their axis. The sum of the transducer voltages was amplified 
and digitized at 500 Hz. Since the force transducers are only sensitive to one direction of stress, we modified the arrangement of the transducers in order to measure the two perpendicular forces, lift and drag, in different runs of the experiment.  

The action of the plow pushes a triangular mound of sand in front of the blade.  The geometry and flow of this plowed material is crucial to producing the lift and drag forces.  We measured the upper position of this mound on the blade using a one dimensional laser position sensor (optoNCDT 1302).
%
 This device provided the length $\ell$ of the part of the mound that was in contact with the plow blade, as shown in Fig.~\ref{mound}.  The length $\ell$ was measured to within an accuracy of 0.2 mm.  
 
 Further measurements were performed using a laser sheet which allows for the computation of the entire shape of the mound of plowed sand. 
  We found that this shape can be reasonably approximated by a triangular prism with a constant dynamic angle of avalanche $\theta \simeq 35^{\circ}$.
 
  The mound was uniform across the front of the plow, which had lateral width $w$, and the sand was prevented from escaping around the ends of the plow by thin forward facing fins on each end of the plow.  Given the angle of attack of the plow $\alpha$, the volume $V$ of the mound is determined if $\ell$ and the angle of avalanche $\theta$ are known.  The mound contains sand with mass $M = \rho \phi V$, where $\rho$ is the density and $\phi$ is the compaction of the grains.   
%
Combining this information gives the mass of the sand in the mound
 \begin{equation}
M= \rho \phi V = \frac{1}{2} w \, \rho \, \phi\, \ell^2\sin^2{\alpha}\left(\frac{1}{\tan{\alpha}}+\frac{1}{\tan{\theta}}\right)~.
\label{eq_Vsand}
\end{equation}

\begin{figure}
\includegraphics[width=7cm]{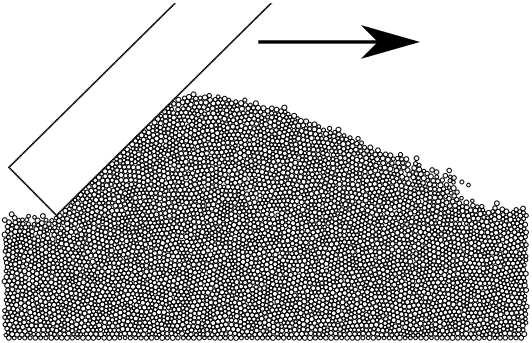}
\caption{A typical snapshot of the 2D molecular dynamics simulations (just showing the region of interest). A plow is dragged at constant velocity and vertical position over a layer of grains which is periodic in the horizontal direction.}
\label{DEM}
\end{figure}

 In this experiment, $w=15~\rm{cm}$ and $\rho=2500~\rm{kg.m^{-3}}$.  We will show in what follows that the length $\ell$ changes with the speed $v$ and position $y$ of the plow, while the other quantities in Eqn.~\ref{eq_Vsand} remain constant.

 The forward flow down the front of the mound produced a constant angle of avalanche which was found to be $\theta=35^{\circ}$ for any plow velocity $v$ and angle of attack $\alpha$.  
  By collecting and weighing the mound of plowed sand we found the compaction of the mound was essentially constant with $\phi=0.45$.  
 It will emerge that the mass $M$ is the main dynamically important quantity that is required to understand the lift and drag forces on the plow blade.
 
 %
 %
 %
 

The experimental protocol was as follows.  Initially, the plow was lifted above the sand surface and remained  empty.   Then its vertical position $y$  was slowly decreased until it plowed ahead of it a mound of grains with a mass of nearly 1~kg. The plow was then kept fixed at this position for at least 10 rotations around the track. After this pre-conditioning step, force measurements were begun. The vertical position of the plow was increased in steps of approximately 10 $\mu$m every 2 seconds. In this way, the mound ahead of the plow was slowly drained, so that after 5 to 15 rotations it was empty again. Lifting the plow five times faster or slower did not change the measured forces significantly, so we may assume that the system evolved quasi-statically.

\section{Numerical simulation \label{DEM_simulation}}


Two-dimensional molecular dynamics simulations were carried out to model the granular motion in the vertical plane perpendicular to the face of the plow. A snapshot of such a simulation is shown in Fig.~\ref{DEM}.  While such 2D simulations cannot provide quantitative agreement with experiments, they provide good qualitative agreement and can be used to gain insight into the origin of the lift and drag forces.  In addition, the simulations allow studies of the positions, forces, velocities and stresses on individual simulated grains, which are difficult to measure directly in an experiment.


The simulation models the individual grains as deformable disks, rotating and colliding with one another. The collision forces acting on each grain are computed each time step and the equations of motion are integrated using the Verlet method~\cite{both01}. The collision force acting between two colliding grains is computed from their overlap $\delta$. A spring-dashpot scheme is used to compute the normal force, $F_n$, given by
\begin{equation}
F_n =  2 \frac{k  r}{R} \delta + \frac{\eta R }{2 r} \dot{\delta}~,
\label{normal}
\end{equation}
where $R$ is the mean radius of the grains and $1/r = 1/r_i +1/r_j$, where $r_i$ and $r_j$ are the radii of the two colliding grains, and $k$ and $\eta$ are parameters.  The first term describes a Hertz's law repulsion due to the small overlap $\delta$ of the two disks (whereas $F_n \propto \delta^{3/2}$ for spheres).
The second term in Eqn.~\ref{normal} describes the dissipation during collisions, which is linear in the velocity $\dot{\delta} = d {\delta}/dt$.

\begin{figure}
\includegraphics[width=8cm]{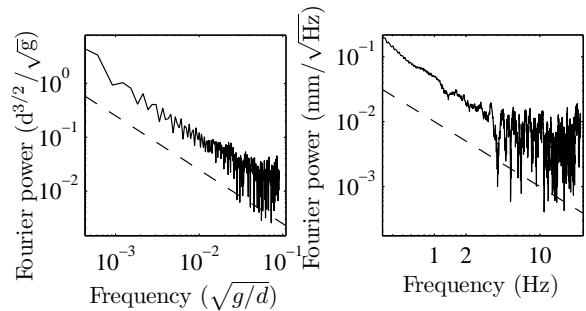} 
\caption{The Fourier power spectrum of a time series of the lift force $F_L$ found in experiments (left panel) and simulation (right panel).  The spectrum of the drag force $F_D$ is similar.  No special frequencies are evident in the flow in either case. The dashed lines have a slope -1. }
\label{FFT}
\end{figure}

The tangential force acting between colliding grains, $F_t$, was computed using a two-parameter regularized Coulomb scheme~\cite{regularized}:

\begin{equation} |F_t| = \min(\mu F_n, \gamma_t v_s)~,
\label{coulomb_scheme}
\end{equation}
where the microscopic friction coefficient $\mu$ was 0.3, while the slope of the regularized region was $\gamma_t=100$ and $v_s$ is the sliding velocity of the contact.
A tangential spring model~\cite{regularized} was also tested and showed no significant differences. 


\begin{figure}[htb]
\includegraphics{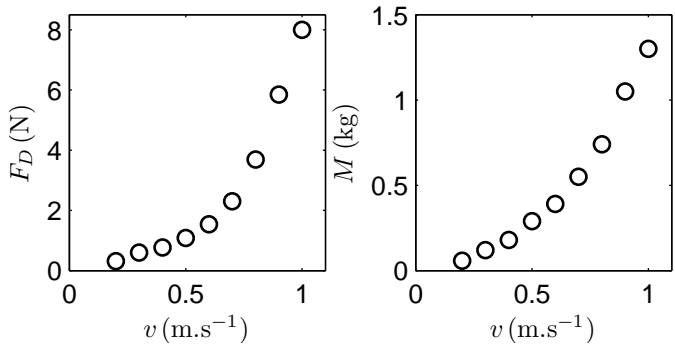}
\caption{On the left, $F_D$ as a function of $v$ for a constant altitude of the bottom of the plow. On the rigth, $M$ as a function of $v$.}
\label{drag_mass_vs_v}
\end{figure}

\begin{figure}[htb]
\includegraphics[width=8cm]{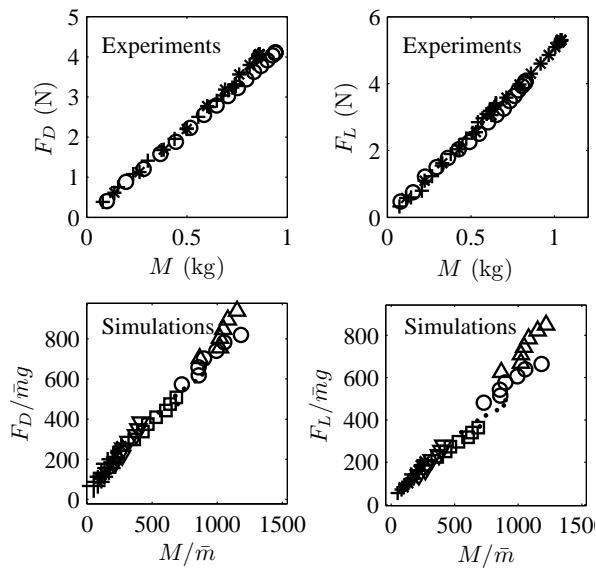} 
\caption{The drag and lift forces $F_L$ and $F_D$ as functions of the plowed mass $M$,  for different velocities at fixed $\alpha$=45$^\circ$. $\bar{m}$ is the average mass of the grains in the simulation.    For the experiments, the symbols are: $+$ $v$=0.1 m s$^{-1}$, $*$ $v$=0.5 m s$^{-1}$ and $\circ$ $v$=1 m.s$^{-1}$.  For the simulation results,  the symbols are: $+$ $v=2 \sqrt{gd}$, $\triangledown$ $v=3 \sqrt{gd}$ , $\square$ $v=4 \sqrt{gd}$, $.$ $v=5 \sqrt{gd}$, $\circ$ $v=6 \sqrt{gd}$, $\triangle$ $v=7 \sqrt{gd}$. }
\label{rescale} 
\end{figure}

A typical numerical simulation used 20~000 disks with a 20\% polydispersity in their diameter to prevent crystallization, giving a periodic domain of length $L = 500\,d$ and depth  $y_0 = 40\,d$.  We used the average diameter $d$, the average mass ${\bar m}$ and $\sqrt{d/g}$ as the units of length, mass and time respectively.  In these units, the moment of inertia $I= \frac{ {\bar m} d^2 }{8} = 1/8$.  Unless otherwise specified, we used parameters $k=10^4$ and $\eta=7.085$, so that the coefficient of restitution of a collision is $e=0.8$.   The time step was chosen to be about 1\% of the collision time $\tau=\pi\sqrt{{\bar m}/(2k)}$. 


The simulation used periodic boundary conditions in the $x$ direction, which mimics the circular track of the experiment.  The simulated plow was formed of smaller disks (10 times smaller) fused together.
  The size of the disks forming the plow does not change the results as long as they remain much smaller than the disks in the bed. 
    The layer of disks was prepared by dropping grains with a random initial position and velocity, and then left to settle under gravity.   
    A numerical run began with the plow above the surface.  Its horizontal speed $v$ was constant throughout the run, while its vertical position was slowly decreased through $y_0$ until it reached the desired height $y$.  Thereafter, the numerical system was allowed to evolve to a statistically stationary state (after several ``rotations'').  The lift and drag forces on the plow were defined to be the sum of the vertical and horizontal forces, respectively, on all the grains making up the plow.

 In the 2D simulation, the mass of plowed mound can be found from a similar geometrical measurement of $\ell$ following Eqn~\ref{eq_Vsand}, or by estimating the area occupied by the disks above the level $y$ of the tip of the plow. Since the velocities of the individual grains are known in the simulation, a third definition of $M$ is the following:

\begin{equation}
M  =  \frac{1}{v}{\sum_{i=1}^{N}{m_i v_i}}~,
\end{equation}
which implicitly computes the average mass of the grains which are carried along at the speed of the plow.  
All three methods gave the same results for the simulations.


\begin{figure}
\includegraphics[width=8cm]{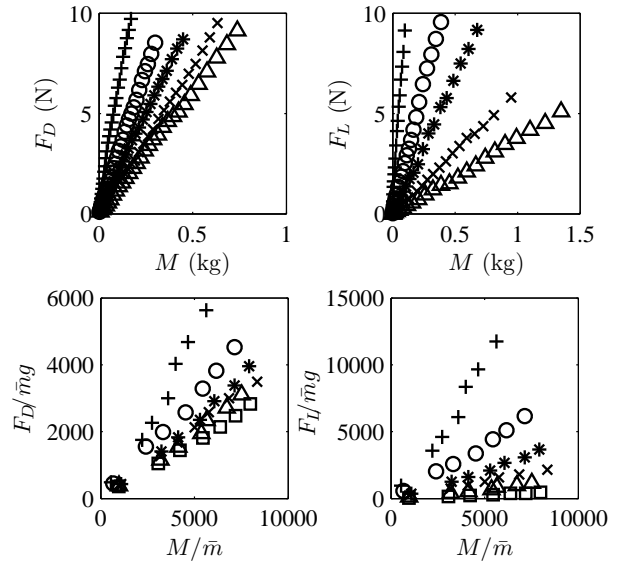} 
\caption{The drag and lift forces $F_L$ and $F_D$ as function of the plowed mass $M$,  for  various angles of attack $\alpha$. The symbols are: $+$ $\alpha$=15$^{\circ}$, $\circ$ $\alpha$=30$^{\circ}$,$*$ $\alpha$=45$^{\circ}$, $\times$ $\alpha$=60$^{\circ}$,$\triangle$ $\alpha$=75$^{\circ}$, $\square$ $\alpha$=90$^{\circ}$ (simulation only).}
\label{angle} 
\end{figure}

 \section{Results}
 \label{results}

In this section, we diagnose the origin and parameter dependence of the lift and drag forces by moving back and forth between the experiment and the 2D simulations.  We will then show how the data in each case may be scaled to produce a general result which can be seen as a simple Coulomb's law of friction with an effective friction coefficient.

\subsection{Frequency spectrum of the lift and drag forces}

We studied time series of the lift and drag forces in both the experiment and in simulation.   Both forces naturally fluctuate as the noisy granular flow proceeds, even in the steady state regime.   As shown in Fig.~\ref{FFT}, the Fourier power spectra of these time series show a roughly power-law dependence on frequency, with no special frequencies evident and an exponent of roughly -1.  This indicates that the flow of the grains is continuous in the regime of plowing speeds we consider, in contrast to the oscillatory flows observed in previous experiments at lower speeds~\cite{Gravish2010}. The qualitative agreement between the experiments and the numerics is excellent.

This rather simple result has important implications for the mechanism of the instability to washboard road, in the case when the plow is free to move vertically.  The absence of special frequencies for flow around the fixed plow indicates that the washboard instability is not merely triggered by some internal oscillatory avalanching mode characteristic of the plowed material alone. Instead, the continuous flow regime must become unstable to an oscillatory mode with a new frequency that emerges from the coupled motion of the grains and the free plow itself.   For the present fixed plow study, this result simply means that we can characterize the lift and drag forces by their average values in the steady state.


\begin{figure*}[htb]
\includegraphics[width=2\columnwidth]{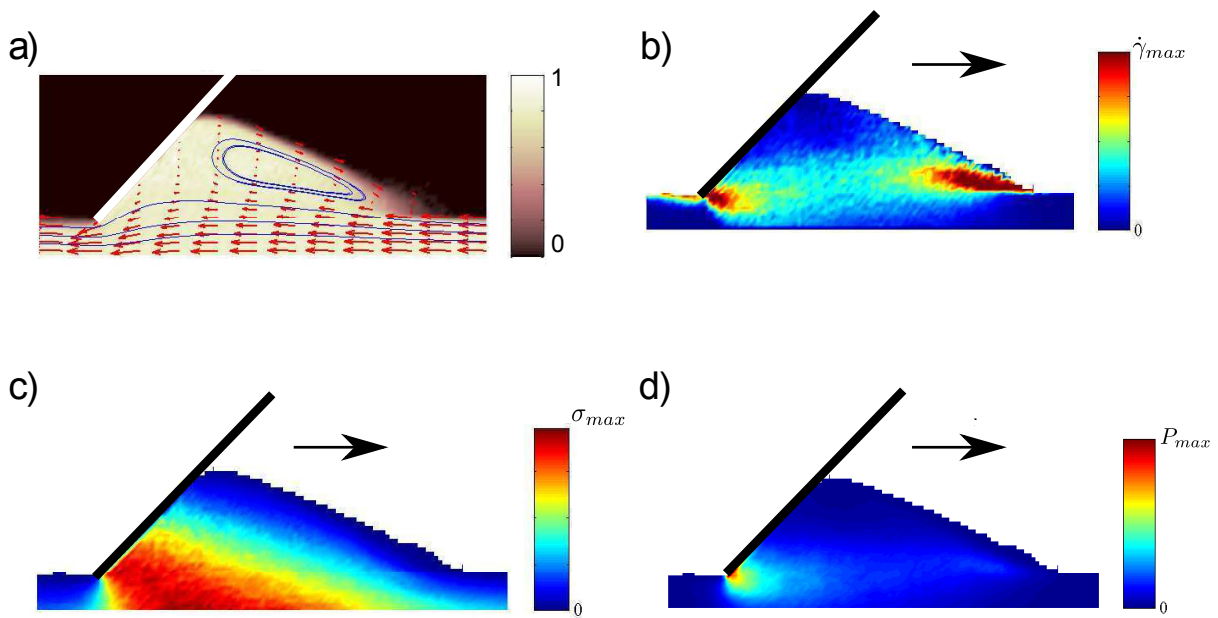}
\caption{Four views of the simulation, showing different aspects. (a) shows the packing fraction $\phi$ (color) and average motion of the grains in the plowed region.  Arrows show the velocity field in the frame of reference of the plow, while the solid lines are streamlines. (b) shows the shear rate $\dot{\gamma}$ in the plowed material, and (c) shows the pressure $\sigma$ distribution.  Finally, (d) shows the dissipated power per unit volume.}
\label{sim_snaps}
\end{figure*}

\subsection{Influence of the plowing speed}

The lift and drag forces increase with the plowing speed $v$ for fixed altitude $y$ and angle of attack, both in the experiment (figure~\ref{drag_mass_vs_v} left) and in the simulations (data not shown).  This increase is a manifestation of the increase in the volume of plowed material that builds up in front of the blade, which increases with $v$ (figure~\ref{drag_mass_vs_v} right).


Fig.~\ref{rescale} shows the remarkable result of this analysis (for an angle of attack of $\alpha = 45^\circ$):
the lift and drag forces exhibit a simple linear relationship when plotted as a function of the plowed mass $M$, roughly independent of the plowing speed $v$.  This shows that $F_L$ and $F_D$ are not directly velocity dependent, as would be the case for hydrodynamic forces, but rather depend only indirectly on velocity, {\it via} the mass of the mound of plowed material and small changes to the effective friction.  Again, note the excellent qualitative agreements between experimental and numerical results.

This behavior is quite consistent with previous experiments on totally immersed intruders~\cite{Wieghardt1975,recent_drag}, in which it was observed that the drag or lift forces have an extremely weak dependence on velocity. This result shows that it is not possible to make any straightforward analogy with a hydrodynamic system such as skipping stones~\cite{Hewitt2011,Clanet2004}, since both viscous and inertial hydrodynamic forces crucially depend on $v$.

The simulations can be extended to very high velocities where the bed becomes fluidized and the drag and lift force become strongly velocity dependent.  In this regime, the simple velocity scaling breaks down and a viscous or turbulent drag model might become relevant. This regime is far above what is accessible experimentally, however.

\subsection{Influence of the angle of attack}


 We repeated the experimental protocol described above for different angles of attack $\alpha$. For each angle, the plow was operated at three different velocities, 0.2~$\rm{ms^{-1}}$, 0.5~$\rm{ms^{-1}}$ and 1.0~$\rm{ms^{-1}}$ and various values of the vertical position $y$. As before, we found that the lift and drag forces depend on the mass of the plowed material $M$ and not directly on the velocity.  Fig.~\ref{angle} shows how the forces depend on $M$ and $\alpha$.  In Fig.~\ref{angle}, the forces were averaged over the three velocities.  A similar protocol was used in the simulations.  As might be expected, the lift force $F_L$ decreases with the angle of attack.  Interestingly, the drag force has a similar dependence on $\alpha$.  These data incorporate some small contributions due to the non-zero thickness of the plow blade.
 The experiments and the simulations show again a very good agreement.
 
 In the following, we will show how this phenomenology can be understood and these data collapsed onto a single curve.  We next turn to the simulations for insight into the interior of the flowing granular material.
 
 
%

\subsection{The interior of the plowed material}

Fig.~\ref {sim_snaps} show several views of the results of the 2D simulation.  We can clearly see the triangular region of the plowed material and the forces and flows within it.  

Within the plowed material, the streamlines in Fig.~\ref{sim_snaps}a, computed in the moving frame of reference of the plow, show a region of forward circulating flow just below the free surface of the plowed material.  Nearer the bottom tip of the plow, the flow is downward. On the face of the plow, the flow velocity is nearly zero relative to the plow.  The packing fraction $\phi$ is quite homogeneous in the plowed sand, similarly to what was measured experimentally, although the values are much higher in this 2D system.  The plowed material slips over the bed at a well-defined region of high shear running horizontally from the bottom of the plow to the end of the free surface.  

This {\it shear band}, shown in Fig.~\ref{sim_snaps}b, cleanly separates the plowed material, which is mostly carried along with the plow, from the sand in the bed, which makes its way under the tip of the plow.  Along this band, there are two localized regions of very high shear, one near the tip of the blade, and the other at the forward toe of the slip face.

\begin{figure}[htb]
\includegraphics[width=6cm]{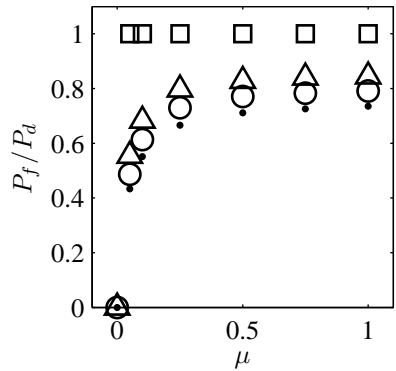}
\caption{The fraction of the total dissipation due to friction, $P_{\rm f}/P_{\rm d}$, as a function of the friction  coefficient $\mu$  for different restitution coefficients $e$.  The symbols correspond to restitution coefficients of: $\square$ e=1 ,$\triangle$ e=0.9, $\circ$ e=0.8, $.$ e=0.5 .}
\label{repartition}
\end{figure}

Using the simulation, we can learn more about the origin of the drag forces by considering the energetics of the flow both locally and globally.  Denoting by $E$, $K$ and $U$ the total energy, total kinetic energy and total gravitational potential energy of the grains, global energy conservation requires
\begin{equation}
\frac{dE}{dt} = \frac{dK}{dt} + \frac{dU}{dt} = P_{\rm i} + P_{\rm d}~,
\end{equation}
where $P_{\rm i} $ is the total power injected by the machinery driving the plow and $P_{\rm d}$ is the total power dissipated by the grains.  In steady state, the total energy is constant and:
\begin{equation}
P_{\rm d} = - P_{\rm i} = F_D v.
\end{equation}
This implies that the drag force is due to the dissipation in the whole of the flowing granular material.
By measuring the drag force and independently integrating the {\it local} dissipation over the flow we checked that  this steady-state result is indeed true.
 In particular this shows that the time step used in the simulation is not too large.

The local dissipation comes from two sources, the collisional restitution and the friction between the grains.  The collisional dissipation is due to the form taken for the normal forces, given by Eqn.~\ref{normal} and depends on the parameters $k$ and $\eta$, which contribute to the coefficient of restitution $e$.  The frictional dissipation depends on the form of the tangential forces and depends on the friction coefficient $\mu$ used in the regularized Coulomb scheme given by Eqn.~\ref{coulomb_scheme}. It is possible to calculate the total power dissipated by each of these mechanisms, so that
\begin{equation}
P_{\rm d} = P_{\rm c} (e) + P_{\rm f} (\mu) ~,
\end{equation}
where $P_{\rm c}$ and $P_{\rm f}$ are the total dissipation by collisional and friction forces, respectively.  Fig.~\ref{repartition} shows the fraction of the total dissipation due to friction forces as a function of the friction coefficient $\mu$ for various values of $e$. As long as the friction coefficient is reasonably high ($\mu>0.2 \simeq \tan 11^{\circ} $), most of the dissipation is frictional (approximately 80\%), even for relatively low values of the restitution coefficient ($e=0.5$). For lower values of the friction it is expected that most of the total dissipation should originate from the inelasticity of the collisions.

Returning to Fig.~\ref{sim_snaps}, we can use the simulation results to examine the localization of the dissipation within the flow.  Fig.~\ref{sim_snaps}c shows that the pressure is concentrated near the tip of the plow.  This pressure maximum is much larger than what the hydrostatic pressure would be at that depth.  The combination of the high pressure with the high rate of shear at the tip of the plow gives a highly localized region of dissipation, as shown in Fig.~\ref{sim_snaps}d.  This dissipation is due to the large loading of the sliding frictional contacts in this small region.

The simulation thus gives us a rather clear picture of the processes within the flow that give rise to the dissipation, and hence the drag force.  The plowed material is an undilated region of circulating flow with a well defined triangular shape, as sketched in Fig.~\ref{mound}.  This material slides over the grains in the bed at its lower surface.  Its geometry and mass $M$ depend on the plowing speed $v$, plow depth $y$ and angle of attack $\alpha$ in such a way that the lift and drag forces are linear in $M$ and independent of $v$.

\section{Discussion}
\label{discussion}

\begin{figure}
\begin{center}
\includegraphics[scale=0.7]{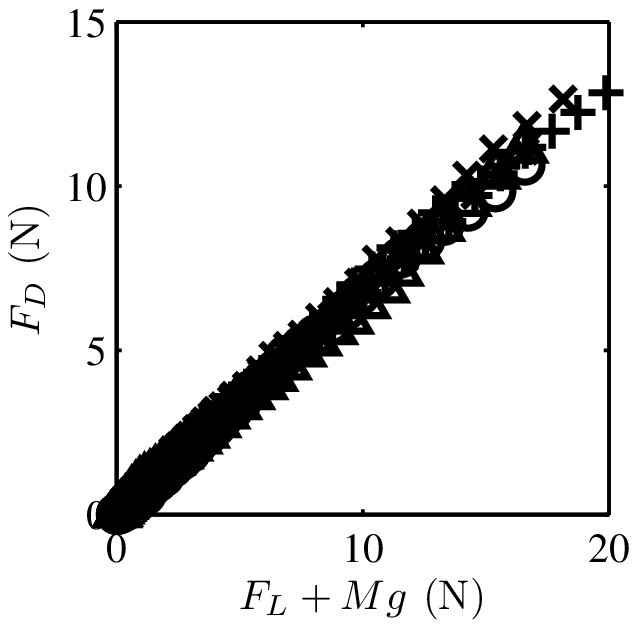}
\includegraphics[scale=0.7]{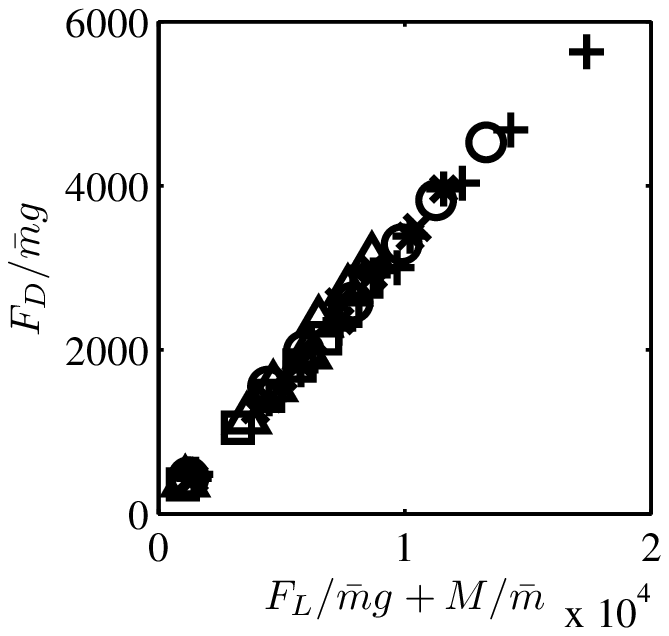} 
\end{center}
\caption{ The drag force $F_D$ as a function of total downward force $F_L+M g$ on the plowed material, modeled as a simple sliding block.   Good data collapse is found for all speeds $v$, vertical positions $y$ and angles of attack $\alpha$, in both the experiment (upper panel) and simulation (lower panel).}
\label{mu_eff}
\end{figure}

The linear relationship between the lift and drag forces and the plowed mass $M$ suggest that all the forces might be accounted for by modeling the plowed material as a solid sliding block moving over a flat surface with Coulomb friction. The lift and drag forces {\it on the plow}, shown in Fig.~\ref{mound}, have corresponding equal and opposite reaction forces {\it on the plowed material}, considered as a triangular solid block.  The reaction to the lift force $F_L$ acts downward on the block, while the reaction partner of the drag force $F_D$  pushes the block forward in the direction of $\vec{v}$.  In addition, the block weight $Mg$, is exerted downward.  Modeling the shear band shown in Fig.~\ref{sim_snaps}b as a simple sliding surface with Coulomb friction subjected to a normal force $F_L + Mg$, we arrive at a simple relationship between $F_D$, $F_L$ and $M$,
 \begin{equation}
F_D = \mu_{\rm eff}  ( F_L+M  g )~.
\end{equation}
Fig.~\ref{mu_eff} shows that this model relationship achieves a near perfect collapse of all the data for both experiment and simulation, using a single universal value of $\mu_{\rm eff}$ in each case.  For experiments, we find  $\mu_{\rm eff} = 0.7$, while the simulations give  $\mu_{\rm eff} = 0.3$.  This difference reflects the rather ideal nature of the 2D simulations, that used perfect disks, which makes them only a very qualitative model of the real grains.  Nevertheless, the main features of the flow are recovered by the simulation, including all the qualitative parameter dependences observed experimentally. 
This universal effective friction law demonstrates that the horizontal drag force $F_D$ is a simple function of the vertical force $F_L+M g$ and depends only indirectly on the altitude $y$, the velocity $v$ and the angle of attack $\alpha$. 



%
%


\section{Conclusion \label{conclusion}}

We have studied both experimentally and numerically the drag and lift forces on an inclined plow blade acting on the surface of a dry granular material.  We considered the case of a wide plow which had no grains flowing around its ends. The flow was nearly two-dimensional and could thus be simulated in the vertical plane.  We studied how the forces depended on the mass and geometry of the mound of granular material transported by the blade.  Using molecular dynamics simulation, we examined the forces, flow and energy dissipation within this plowed material.  We found that the flow is steady in the velocity regime we studied, so that the washboarding instability of a free plow could not be explained by any pre-existing unsteady motion within the plowed material in front of the fixed plow. The lift and drag forces did not depend ,significantly on the velocity of the plow if they are considered as functions of the mass of the transported material.  We also found that the two forces and the weight of the plowed material could be combined into a simple relationship in which the plowed material behaves as a solid block sliding over the underlying granular bed.  The sliding was characterized by a single effective Coulomb friction coefficient.  

The case of the fixed plow discussed in this paper must be generalized in order to establish a linear stability analysis of a free plow, which is unstable to the formation of washboard ripples.  Non-stationary states of the plow and of the mound of plowed material, and the time-dependent effect of these on the lift and drag forces, must be accounted for in such a stability analysis. In future work the response of the system to small, imposed vertical oscillations at various frequencies will be studied using the same experimental and simulation techniques described here. This approach will shed some light on the dynamic origin of the washboard instability.



\begin{acknowledgments}
We thank Bruno Andreotti, Nick Gravish, Paul Umbanhowar and Daniel Goldman for helpful discussions.  This research was supported in part by the Natural Science and Engineering Research Council (NSERC) of Canada.
\end{acknowledgments}

\end{document}